\documentclass[aps,pra,showpacs,twocolumn,superscriptaddress]{revtex4}

\usepackage{graphicx,color}
\usepackage{amsmath,amsthm,amsfonts,amssymb,bm}
\usepackage{times}
\usepackage{epsf}

\usepackage[T1]{fontenc}
\usepackage[utf8]{inputenc}
\usepackage{ulem}

\newcommand{\ket}[1]{\left| #1 \right\rangle}

\newcommand{\ketbra}[2]{\left|#1\middle\rangle\middle\langle#2\right|}

\usepackage{graphicx}
\begin{document}

\title{Quantum Correlations and Coherence in Spin-1 Heisenberg Chains}

\author{A. L. Malvezzi}
\affiliation{Faculdade de Ci\^encias, UNESP - Universidade Estadual Paulista, Bauru, SP,
17033-360, Brazil}
\author{G. Karpat}
\affiliation{Turku Center for Quantum Physics, Department of Physics and Astronomy,
University of Turku, FI-20014,
Turun yliopisto, Finland}
\affiliation{Faculdade de Ci\^encias, UNESP - Universidade Estadual Paulista, Bauru,
SP, 17033-360, Brazil}
\author{B. \c{C}akmak}
\affiliation{Department of Physics, Ko\c{c} University, \.{I}stanbul, Sar\i yer 34450, Turkey}
\affiliation{Instituto de Fisica Gleb Wataghin, Universidade Estadual de Campinas, P.O.
Box 6165, Campinas, SP, 13083-970, Brazil}
\author{F. F. Fanchini}
\email{fanchini@fc.unesp.br}
\affiliation{Faculdade de Ci\^encias, UNESP - Universidade Estadual Paulista, Bauru,
SP, 17033-360, Brazil}
\author{T. Debarba}
\affiliation{Universidade Tecnol\'ogica Federal do Paran\'a (UTFPR), Campus Corn\'elio Proc\'opio.
R. Alberto Carazzai, 1640, Corn\'elio Proc\'opio, PR, 86300-000 - Brazil.}
\author{R. O. Vianna}
\affiliation{Departamento de Fisica - ICEx - Universidade Federal de Minas Gerais,
Av. Pres. Antonio Carlos
6627, Belo Horizonte, 31270-901, Brazil}

\begin{abstract}
We explore quantum and classical correlations along with coherence in the ground states of spin-1 Heisenberg
chains, namely the one-dimensional XXZ model and the one-dimensional bilinear biquadratic model, with the
techniques of density matrix renormalization group theory. Exploiting the tools of quantum information theory,
that is, by studying quantum discord, quantum mutual information and three recently introduced coherence measures
in the reduced density matrix of two nearest neighbor spins in the bulk, we investigate the quantum phase
transitions and special symmetry points in these models. We point out the relative strengths and weaknesses
of correlation and coherence measures as figures of merit to witness the quantum phase transitions and symmetry
points in the considered spin-1 Heisenberg chains. In particular, we demonstrate that as none of the studied
measures can detect the infinite order Kosterlitz-Thouless transition in the XXZ model, they appear to be able
to signal the existence of the same type of transition in the biliear biquadratic model. However, we argue that
what is actually detected by the measures here is the SU(3) symmetry point of the model rather than the
infinite order quantum phase transition. Moreover, we show in the XXZ model that examining even single site
coherence can be sufficient to spotlight the second-order phase transition and the SU(2) symmetry point.
\end{abstract}


\maketitle

\section{Introduction}

The investigation of many-body quantum systems have revealed very interesting and deep physical
concepts such as quantum phase transitions (QPT). QPTs are abrupt changes in the ground state of a
quantum many-body system as one or more parameters of the Hamiltonian is varied at absolute zero
temperature \cite{qpt}. In contrast to thermal phase transitions, which are driven by thermal fluctuations in
the system, QPTs are driven by quantum fluctuations stemming from the uncertainty principle. However,
it is also possible to see the effects of a QPT at sufficiently low but finite temperatures where the
quantum fluctuations are not washed away by the thermal effects. Traditionally, phase transitions are
classified based on the non-analytic behavior in the derivatives of the ground state energy.
In particular, a discontinuity in the first derivative of the ground state energy signals a first order
transition. On the other hand, a discontinuity or divergence in the second derivative of the ground state
energy is recognized as a second order transition in which case the transition is associated with a symmetry
breaking. A more involved type of phase transition, which does not fit to the traditional
classification scheme, is known as the Kosterlitz-Thouless (KT) transition. In this case, there is no
divergence or discontinuity in the derivatives of the ground state energy and no symmetry breaking thus
KT transitions are said to be of infinite order \cite{contqpt}.

Quantum many-body systems possess correlations of various different nature due to the interaction
among their constituents. Therefore, in addition to the traditional ways of witnessing quantum phase
transitions, it has been recently suggested that the tools of quantum information theory \cite{qit}
can also be exploited to characterize the transition points (TPs) of quantum phase transitions.
Especially, in quantum spin models, the behavior of entanglement \cite{entangle}, quantum discord
\cite{discord} and many other correlation measures have been investigated, and their performance in
detecting the TPs of the QPTs have been discussed \cite{allsc,cohsc}. Recently, a new line
of research has emerged that concerns itself with the characterization and quantification of quantum
coherence contained in a quantum state \cite{baumgratz,shao,streltsov,girolami,xi}. Based on these new quantum coherence measures,
similar analysis have been done in the ground states of several spin chains \cite{cohsc}. However, many of
these studies focusing on quantum correlations in spin chains have been done for spin-$1/2$ systems
\cite{allsc,cohsc}, where analytical solutions are available in many cases. On the other hand,
spin-$1$ models have richer phase diagrams and show more complex physical phenomena, yet, methods for
obtaining the ground state of such systems are rather more involved \cite{alcaraz,cao,chen,kitazawa,legeza,liu,orus2010,orus,sakai,su,wang,zhou2008,zhou2008-2,zhou2008-3,dechiara,lepori,power,chubukov}. For instance, a very important distinctive property of the integer-spin quantum systems as compared to the half integer ones is the Haldane conjecture, which states that the system has a gapped ground state, giving rise to the so-called Haldane phase \cite{haldane}.

In this work, we will consider two very well known one-dimensional spin-1 Heisenberg models, namely,
the spin-1 XXZ chain and the spin-1 bilinear biquadratic chain. Both of these models have
been under extensive investigation in the literature from different perspectives due to
the rich physics they exhibit. Here we obtain the ground state of these systems by making use
of the methods of density matrix renormalization group theory (DMRG). Then, we extensively
investigate the behavior of mutual information, quantum discord and three recently introduced
coherence measures namely, relative entropy of coherence, $l_1$ norm of coherence \cite{baumgratz} and Wigner-Yanase skew
information \cite{girolami,WYSI}, for the reduced density matrix of two nearest neighbor spins in the bulk.
Our analysis lets us establish relations between the phase transitions and symmetry points in the
considered spin-1 Heisenberg chains and the studied correlation and coherence measures.

This paper is organized as follows. In Sec. II, we introduce the spin-one Heisenberg models used
in this study along with the DMRG techniques required to obtain the numerical solution of these
models. Section III presents the definitions of the considered correlation and coherence measures.
We present our results in Sec. IV and conclude in Sec. V.

\section{Models}

In the section, we briefly discuss the different phases that the one-dimensional XXZ model and
the one-dimensional bilinear biquadratic model favor with respect to their characteristic parameters,
and the nature of phase transitions occurring among these phases. In order to calculate the ground state
of the model Hamiltonians, we use the standard DMRG infinite system method \cite{dmrg}. In this version of DMRG,
an open chain is grown iteratively by adding two sites at a time to the center of the chain. At each step
the ground state for the whole chain is calculated and a renormalization procedure is performed. Typically,
after a few hundred iterations the two central sites are embedded in a bulk. The reduced density matrix for
this two central sites can then be obtained from the ground state. It is important to stress that, despite
the renormalization, the spin interaction between the two central sites is always kept exact. As the model
parameters vary, truncation errors in the renormalization procedure range from $10^{-10}$ to $10^{-6}$ with
the upper limit occurring around second order phase transitions, where quantum fluctuations are stronger.

\subsection{Spin-1 XXZ Chain}

The Hamiltonian describing the one-dimensional spin-1 XXZ model with nearest neighbor
interaction reads
\begin{equation}
H= \sum_{i=1}^{N} [S_i^x S_{i+1}^x+S_i^y S_{i+1}^y+\Delta S_i^z S_{i+1}^z],
\end{equation}
where $N$ is the total number of sites, $\textbf{S}_i$ denotes the spin-1 operator at
the site $i$, and $\Delta$ characterizes the anisotropy of the spin-exchange interaction
in the model. It is well established that the model have four different phases depending
on the value of the anisotropy parameter \cite{alcaraz,kitazawa,sakai}. The system is in a ferromagnetic phase when
$\Delta<-1$. There is a first-order phase transition at the TP $\Delta_{c1}=-1$, which
separates the ferromagnetic phase from the $XY$ phase. At the second TP $\Delta_{c2}$,
the system exhibits an infinite order phase transition (that is believed to be of KT type),
from the $XY$ phase to the Haldane phase, which extends over the region  $\Delta_{c2}<\Delta<\Delta_{c3}$.
There is also a second-order phase transition taking place at the TP $\Delta_{c3}$
from the Haldane phase to the N\'{e}el phase, belonging to the two-dimensional Ising
universality class. Even though the exact values of both TPs $\Delta_{c2}$
and $\Delta_{c3}$ have been the subject of various numerical studies, it is widely accepted that
XY-Haldane and Haldane-N\'{e}el transitions respectively occur at the TPs $\Delta_{c2}\approx0$
and $\Delta_{c3}\approx1.185$ \cite{kitazawa,sakai}. It should also be emphasized that, in addition to the phase
transition points, the model also has a particular $SU(2)$ symmetry point at $\Delta=1$.

\subsection{Spin-1 Bilinear Biquadratic Chain}

The Hamiltonian of the one-dimensional spin-1 bilinear biqudaratic chain can be written as
\begin{equation}
H=\sum_{i=1}^N [\cos\theta (\textbf{S}_i \cdot \textbf{S}_{i+1})+
\sin\theta (\textbf{S}_i \cdot \textbf{S}_{i+1})^2],
\end{equation}
where $N$ is the total number of sites, $\textbf{S}_i$ denotes the spin-1 operator
at the site $i$, and $\theta\in[0,2\pi)$ is the angle quantifying the amount of coupling
between nearest neighbor spins. The model system has an especially rich phase diagram.
In the parameter region $-0.25\pi<\theta<0.25\pi$, the system is in the Haldane phase.
At the TP $\theta_{c1}=0.25\pi$, there is a transition of the KT type, separating the Haldane
phase from the gapless trimerized phase. A first-order transition from the trimerized phase
to the ferromagnetic phase occurs at the TP $\theta_{c2}=0.5\pi$. As the system favors the
ferromagnetic phase throughout the parameter region $0.5\pi<\theta<1.25\pi$, another first-order
transition takes place at the TP $\theta_{c3}=1.25\pi$ from the ferromagnetic phase to
the gapped dimerized phase. Finally, there exists a second-order transition between the
dimerized phase and the Haldane phase at the TP $\theta_{c4}=1.75\pi$.
Although it has been also suggested that the model might exhibit a non-dimerized nematic
phase in the region $5\pi/4<\theta<1.33\pi$ with a KT type transition at $\theta=1.33\pi$ \cite{chubukov},
it has been recently shown that no such nematic phase exists and the system remains in the dimerized phase
all through this region \cite{hu}. It is also worth to mention that at
$\theta=0.1024\pi$ the system corresponds to the Affeck-Kennedy-Lieb-Tasaki (AKLT) model \cite{aklt} with
an exact valence bond ground state, and at $\theta=1.5\pi$ it can be solved exactly by the Bethe
ansatz method \cite{bethe}. Last but not least, we stress that besides being a TP of the model, $\theta=0.25\pi$
is also special in that the system has a $SU(3)$ symmetry \cite{sutherland}.

\section{Correlations and Coherence}

This section serves as a brief introduction to description of the figures of merit that
we will be using throughout this work, i.e., quantum mutual information, quantum discord,
relative entropy of coherence, $l_1$ norm of coherence, and Wigner-Yanase skew information
based measure of coherence.

The relevance of these measures to the study of quantum phase transition follows from different reasons. Quantum discord, for example, was broadly studied in the quantum critical systems and brought various new insights to the field when compared with entanglement measurements. Quantum discord can detect QPTs even when the entanglement measures fail to do so and it can be used even for thermal systems. However, to evaluate it for spin dimensions higher than two spin-1/2 particles is a highly demanding task. To overcome this difficulty, we have introduced a simple numerical procedure that will allow the analysis of quantum discord in high dimensional spin systems. Indeed, in our view, it will serve as an efficient new tool of the quantum information theory for the study of QPT. On the other hand, once again we would like to emphasize that, to the best of our knowledge this is the first work which calculates the recently introduced coherence measures in a spin-1 model. The importance of the calculation of coherence measures stem from the facts that they can be used as a resource in quantum computing protocols \cite{gerardo}, they can be calculated even for single spin density matrices, they are experimentally friendly quantities to calculate and they are analytically (and easily) computable even for high spin dimensions.

\subsection{Quantum Discord}

Let us commence by introducing the quantum mutual information. It quantifies the total
amount of classical and quantum correlations in a bipartite quantum state $\rho_{AB}$ as
\begin{equation}
\mathcal{I}(\rho_{AB})=S(\rho_A)+S(\rho_B)-S(\rho_{AB}),
\end{equation}
where $\rho_{A(B)}$ are the reduced density matrix of subsystem A(B) respectively and
$S(\rho)=-\text{Tr}\{\rho\log_2\rho\}$ is the von Neumann entropy. On the other hand,
the classical correlation, which is the maximum amount of classical information that
can be obtained about the subsystem $A$ by performing local measurements on the
subsystem $B$, is given by
\begin{equation}
\mathcal{C}(\rho_{AB})=\max_{\{\Pi_{k}^{B}\}}\left\{ S\left(\rho_{A}\right)-
\sum_{k}p_{k}S\left(\rho_{A|k} \right)\right\},
\end{equation}
where the operators $\{\Pi_{k}^{B}\}$ constitute a positive operator valued measure (POVM)
acting only on the subsystem $B$, and $\rho_{A|k}=Tr_B(\Pi_{k}^{B}\rho_{AB}\Pi_{k}^{B})/p_k$
is the remaining state of the subsystem $A$ after obtaining the outcome $k$ with probability
$p_k=Tr_{AB}(\Pi_{k}^{B}\rho_{AB}\Pi_{k}^{B})$ in the subsystem $B$. Then, the amount of inaccessible
information, by means of local measurements, defines the quantum discord as \cite{discord}
\begin{equation}
\mathcal{D}(\rho_{AB})=\mathcal{I}(\rho_{AB})-\mathcal{C}(\rho_{AB}).
\end{equation}

In the following, we will describe a numerical recipe to efficiently calculate quantum discord. We should
underline that,unlike the most works in the literature, where quantum discord is evaluated for a pair of
qubits and only using projective measurements, our method makes it possible to calculate quantum discord for
the composite system of two spin-one objects and using POVMs rather than projective measurements. Nonetheless,
we restrict ourselves to the projective measurements in this work for the sake of simplicity.

\subsubsection*{Numerical evaluation of quantum discord}

Although the calculation of quantum discord is an NP-complete problem \cite{huang}, i.e. the necessary time to
obtain the value for it grows exponentially with the Hilbert space dimension, we present
a method to numerically compute quantum discord.  In order to find the global minimum of quantum discord
under projective measurements, we have to search all the Hilbert space to find the optimal orthonormal basis.
One way of doing so is to generate a random unitary matrix, whose eigenvectors are used as a starting point for
a global optimization technique, like steepest descent or variable metric methods as implemented in MATLAB,
for instance. The minimization is repeated for different starting points.

The random unitary matrix is obtained by means of a circular unitary ensemble (CUE), which consists of
all unitary  matrices with Haar measure in the unitary group, following the technique proposed in \cite{randomzyc99,randomzyc}.
The idea is to generate Euler angles ($\phi, \psi, \chi$), such that the arbitrary unitary matrix $U$
be composed from unitary transformations $E^{(i,j)}(\phi, \psi, \chi)$ in two-dimensional subspaces \cite{randomzyc}.
The non-zero elements of the matrices $E^{(i,j)}$ are:
\begin{align}
E^{(m,n)}_{kk} &= 1, \quad k\neq m,n\\ \nonumber
E^{(m,n)}_{mm} & = \cos{\phi_{mn}}e^{i\psi_{mn}}\\ \nonumber
E^{(m,n)}_{mn} & = \sin{\phi_{mn}}e^{i\chi_{mn}} \\ \nonumber
E^{(m,n)}_{nm} & =  - \sin{\phi_{mn}} e^{-i\chi_{mn}} \\ \nonumber
E^{(m,n)}_{nn} & =  \cos{\phi_{mn}} e^{-i\psi_{mn}}.  \nonumber
\end{align}
The matrices of Euler angles $\psi$ and $\phi$ have dimension $(N-1)\times (N-1)$, and  the matrix
$\chi$ has dimension $(N-1)\times 1$,  where $N$ is the dimension of the Hilbert space. The angles
in $\psi$ and $\chi$ must be taken uniformly in the interval $[0,2\pi)$. The angles in  $\phi$ are
given by  $ \arcsin(\xi_{rs}^{1/(2r+2)})$, for $r=0, 1,\dots,N-2$, with  $\xi_{rs}$ uniformly
distributed in the interval $[0,1)$. The random unitary matrix $U$ reads:
\begin{equation}
U = e^{i\alpha} E_1 E_2,\dots E_{N-1},
\end{equation}
where $\alpha$ is also taken uniformly in the interval $[0,2\pi)$, and the matrices $E_{k}$,
for $k=1, 2, \dots, N-1$ read:
\begin{align}\nonumber
E_k =& E^{(N-k, N-k-1)}(\phi_{k-1,k},\psi_{k-1,k},0) \times \cdots \\
\cdots\times & E^{(N,N-1)}(\phi_{0,k},\psi_{0,k},\chi_k).
\end{align}

In our numerical calculations, we generate the angles $\alpha, \phi, \psi$ and $\chi$ by means of
a uniform random vector $x_0$, with $2(N-1)^2+N$ elements. The first element of $x_0$ is the angle
$\alpha$, the next $N-1$ elements correspond to   the vector $\chi$, and the last $2(N-1)^2$ elements
generate  the matrices $\phi$ and $\psi$. Therefore, following the technique presented above, we can
create a random unitary matrix starting point, and search for the global minimum in the space of
unitary matrices with  Haar measure in the unitary group \cite{dyson62}.

With the aid of the Naimark's theorem \cite{Zyc}, the algorithm to optimize the quantum discord under
projective measurements can also be used to perform the optimization under POVMs. Let us review the
procedure. Consider a set of positive semidefinite operators $P_a$, acting on the Hilbert space $\mathcal{X}$,
 of dimension $x$. In order to form a POVM, the $P_a$ must satisfy
\begin{equation}
\sum_a P_a=I_x,
\end{equation}
where $I_x$ is the identity operator acting on $\mathcal{X}$. To each $P_a$, we associate a projector
$\ketbra{a}{a}$, acting on an extended Hilbert space $\mathcal{X} \otimes \mathcal{Y}$. We wish that a von
Neuman measurement in the extended Hilbert space $\mathcal{X} \otimes \mathcal{Y}$, of dimension $x\times y$,
reproduce the statistics of the POVM in the original space $\mathcal{X}$, namely:
\begin{equation}
Tr(P_a \rho)=Tr(A^\dagger \ketbra{a}{a} A \rho),
\end{equation}
where $A$ is an isometry that takes a vector in $\mathcal{X}$ to $\mathcal{Y}$,
\begin{equation}
A^\dagger A = I_x, \,\, A^\dagger \ketbra{a}{a} A= P_a.
\end{equation}
Let $\ket{e_a}$ be the canonical basis in $\mathcal{Y}$, then we choose:
\begin{equation}
\ketbra{a}{a}=I_x \otimes \ketbra{e_a}{e_a}.
\end{equation}
With this choice, the isometry $A$ reads,
\begin{equation}
A=\sum_a \sqrt{P_a}\otimes \ket{e_a}.
\end{equation}
By choosing some arbitrary  ancilla $\ket{u}$ in $\mathcal{Y}$, we can decompose the isometry as:
\begin{equation}
A=U V, \, V=(I_x \otimes \ket{u}),
\end{equation}
where $U$ is a unitary acting on $\mathcal{X} \otimes \mathcal{Y}$.
Finally we have:
\begin{equation}
Tr(P_a \rho)= Tr(Q_a \rho \otimes \ketbra{u}{u}),
\end{equation}
where the projector $Q_a$ reads:
\begin{equation}
Q_a=U^\dagger (I_x \otimes \ketbra{e_a}{e_a}) U.
\end{equation}
We conclude that the POVM $\{P_a\}$ in the original space is equivalent to the projective measurement $\{Q_a\}$ over the state $\rho\otimes \ketbra{u}{u}$
in the extended space. The unitary $U$ is explicitly:
\begin{equation}
U=AV^+,
\end{equation}
where $V^+$ is the pseudo-inverse of $V$.

\subsection{Quantum Coherence}

Although quantum coherence plays a central role in quantum mechanics being a manifestation of the
quantum superposition principle, its quantification has been formalized only very recently.
In particular, a set of conditions that is expected to be satisfied by any proper measure of
coherence has been proposed in Ref. \cite{baumgratz}. Two such measures that we study in this work are
known as the relative entropy of coherence and the $l_1$ norm of coherence. While the former
is defined as
\begin{equation}
C_{\text{re}}(\rho)=S(\rho_{\text{diag}})-S(\rho),
\end{equation}
where $S(\rho_{\text{\text{diag}}})$ is obtained from the state $\rho$ by deleting all of
its off-diagonal elements, the latter is given by the sum of absolutes values
of all off-diagonal elements of $\rho$, that is,
\begin{equation}
C_{l_1}(\rho)=\sum_{i\neq j}|\rho_{i,j}|.
\end{equation}
Naturally, it is only meaningful to talk about coherence measures once we set a specific
basis for incoherent quantum states since coherence is clearly basis dependent.

On the other hand, there is a particular quantity which, despite it does not satisfy \cite{du} the
conditions proposed in Ref. \cite{baumgratz}, can still be considered as a measure of coherence in a
conceptually different way \cite{girolami,streltsov}, i.e., Wigner-Yanase skew information \cite{WYSI}:
\begin{equation} \label{WYSI}
C_{\text{si}}(\rho,K)=-\frac{1}{2}\textmd{Tr}[\sqrt{\rho},K]^2,
\end{equation}
where $K$ is a non-degenerate Hermitian matrix, and $[.,.]$ denotes the commutator. We note
that as the skew information reduces to the variance  $V(\rho,K)=\textmd{Tr}\rho
K^2-(\textmd{Tr} \rho K)^2$ for pure states, it is upper bounded by the variance for
mixed states. In fact, $C_{\text{si}}(\rho,K)$ is a measure of asymmetry
relative to the group of translations generated by the observable $K$, which in turn can be
interpreted as a measure of coherence of the state $\rho$ relative to the eigenbasis of
the observable $K$ \cite{marvian}. From this point on, we will simply refer $C_{\text{si}}(\rho,K)$
as $K$-coherence.

\section{Results}

\begin{figure}[t]
\includegraphics[width=0.55\textwidth]{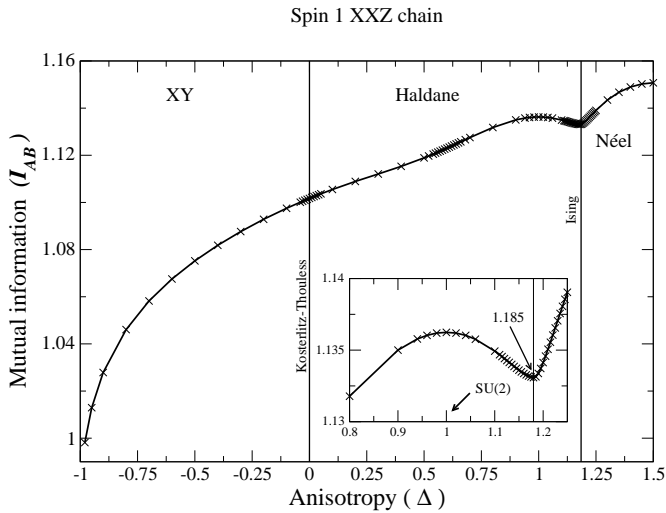}
\caption{Mutual information versus the anisotropy parameter $\Delta$ in the
one-dimensional spin-$1$ XXZ model. Different phases, transition points and the $SU(2)$ symmetry point
are shown.}
\label{fig1}
\end{figure}

In this section, we intend to investigate the behavior of the considered correlation and coherence
measures in the ground state of the spin-$1$ XXZ chain for two nearest neighbor spins in the bulk.
For our purposes, we consider the region where the anisotropy parameter lies in between $-1<\Delta<1.5$.
Observing the Fig. \ref{fig1}, at the TP $\Delta_{c2}\approx0$, where the system has an infinite order phase
transition, we do not notice a non-trivial behavior in the quantum mutual information, i.e., it does not
exhibit either a non-analytical behavior or an extremum. In other words, the mutual information
is not able to detect the existence of the KT type transition in the XXZ chain. On the other hand, at the
point $\Delta_{c3}\approx1.185$, we see a pronounced local minimum, which spotlights the second order
transition that the system has between the Haldane and N\'{e}el phases. Recalling that the XXZ chain also
has a particular $SU(2)$ symmetry point at $\Delta=1$, we can observe that the mutual information shows a
smooth local maximum at this special point.

\begin{figure}[t]
\includegraphics[width=0.55\textwidth]{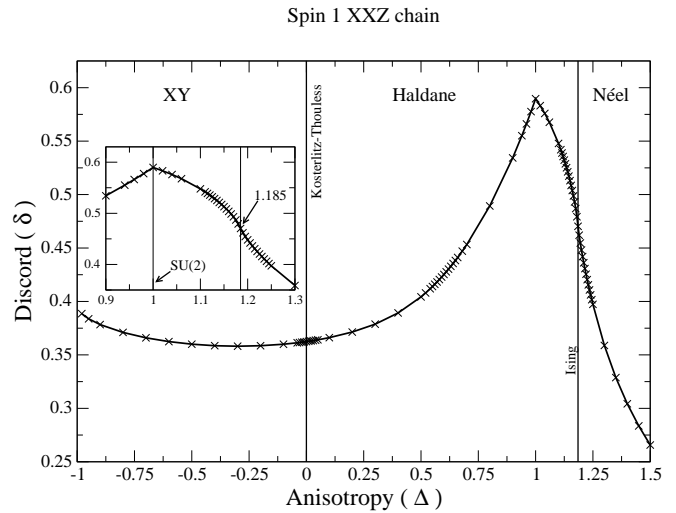}
\caption{Quantum discord versus the anisotropy parameter $\Delta$ in the
one-dimensional spin-$1$ XXZ model. Different phases, transition points and the $SU(2)$ symmetry point
are shown.}
\label{fig2}
\end{figure}

Fig. \ref{fig2} displays the outcomes of our analysis for the quantum discord in the ground state of the XXZ chain.
Similarly to the case of quantum mutual information, quantum discord is not capable of recognizing the
location of the KT type transition. In fact, it is rather expected that neither mutual information nor
quantum discord show a non-analytic behavior at this point, since all derivatives of the ground state energy
and thus the elements of the two-spin density matrix we study are continuous for an infinite order transition.
Moving to the second order transition at the TP $\Delta_{c3}\approx1.185$, we notice that quantum discord
shows an inflection point, that is, the transition point might be easily captured looking at the derivative
of Fig. \ref{fig2} around this point, which would display a quite pronounced minimum. Finally, it is straightforward to
observe that quantum discord has a sharp maximum at the $SU(2)$ symmetry point $\Delta=1$. Indeed, a closer
inspection reveals that quantum discord has a sudden change at this value of the anisotropy parameter. That
is to say that the optimal measurement basis for quantum discord suddenly changes at the $SU(2)$ symmetry
point, resulting in a clear identification of the symmetry point through quantum discord. We note that this
is fundamentally different from the way mutual information feels the existence of the $SU(2)$ symmetry point.

\begin{figure}[t]
\includegraphics[width=0.55\textwidth]{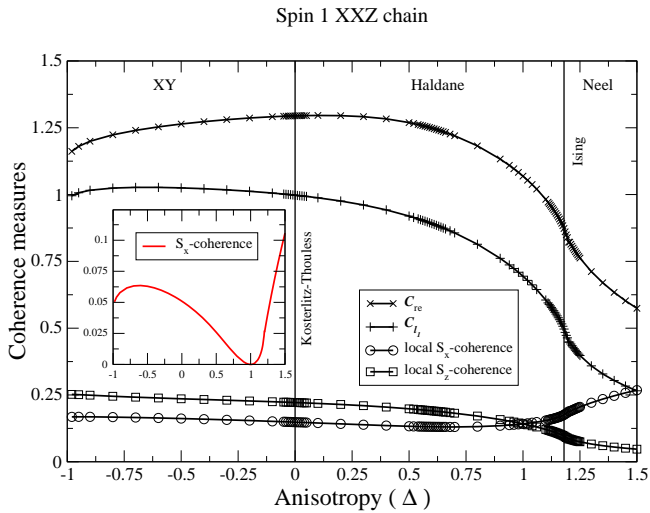}
\caption{(Color online) Quantum coherence measures as quantified by $l_1$ norm of coherence, relative
entropy of coherence and $K$-coherence versus the anisotropy parameter $\Delta$ in the one-dimensional
spin-$1$ XXZ model. Different phases, transition points and the $SU(2)$ symmetry point are shown.}
\label{fig3}
\end{figure}

Next, we explore the quantum coherence using different measures in the ground state of the spin-$1$ XXZ
chain for two nearest neighbor spins in the bulk. In Fig. \ref{fig3}, we plot the relative entropy of
coherence, $l_1$ norm of coherence, the local $S_x$-coherence and the local $S_z$-coherence versus the
anisotropy parameter, where $S_x$ and $S_z$ are the usual spin-1 matrices. Local $K$-coherence means that
the observable $K$ in the definition of $C_{\text{si}}(\rho,K)$ is simply $I \otimes K$. First of all, we
immediately notice that none of the considered coherence measures can spotlight the infinite order KT
transition in the model since they do not exhibit a non-trivial behavior at the transition point.
To put it differently, the KT transition in the spin-1 XXZ model escapes all of the correlation and coherence
measures that we use in our investigation. On the other hand, all four coherence measures can detect the
second order transition occurring at the TP $\Delta_{c3}\approx1.185$ via an inflection point. Turning our
attention to the $SU(2)$ symmetry point in the XXZ model, we observe that the relative entropy of coherence
and $l_1$ norm of coherence are not able to feel the existence of this point at $\Delta=1$. In addition, it
is also not possible to detect the $SU(2)$ symmetry point just by checking the local $S_x$-coherence or the
local $S_z$-coherence individually. However, it is interesting that plotting the $K$-coherence for two
different observables reveals the $SU(2)$ symmetry point through the intersection of these two curves.
That is, the curves of the local $S_x$-coherence or the local $S_z$-coherence intersect at the $SU(2)$
point $\Delta=1$, i.e., the system has the same $K$-coherence at the symmetry point, independently of
the observable $S_x$ and $S_z$.

Lastly, we study the $S_x$-coherence in the ground state of the spin-$1$ XXZ chain for a single spin
in the bulk. The results of this investigation are shown in the inset of Fig. 3. Interestingly, an inflection
point still appears even in the level of single-site coherence at the TP of the second order phase transition
at $\Delta_{c3}\approx1.185$. Furthermore, the $S_x$-coherence vanishes only at the $SU(2)$ symmetry point
$\Delta=1$, pinpointing its location. The reason we do not display the relative entropy of coherence and
$l_1$ norm of coherence here is that they are zero for all values of the anisotropy parameter due to the fact
that the single spin density matrix  is diagonal in $S_z$ basis.

\begin{figure}[t]
\includegraphics[width=0.55\textwidth]{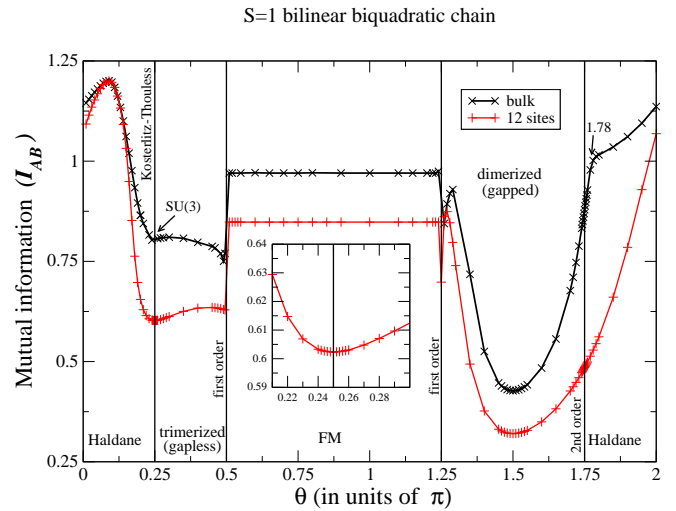}
\caption{(Color online) Mutual information versus the parameter $\theta$ in the one-dimensional spin-$1$
bilinear biquadratic model. Different phases, transition points and the $SU(3)$ symmetry point are shown.}
\label{fig4}
\end{figure}

\begin{figure}[t]
\includegraphics[width=0.55\textwidth]{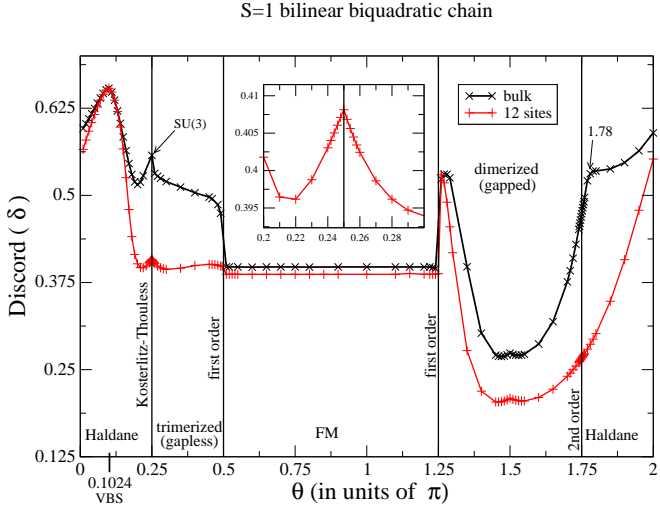}
\caption{(Color online) Quantum discord versus the parameter $\theta$ in the one-dimensional spin-$1$
bilinear biquadratic model. Different phases, transition points and the $SU(3)$ symmetry point are shown.}
\label{fig5}
\end{figure}

Having discussed the correlations and coherence in the spin-1 XXZ chain, we now examine the spin-1 bilinear
biquadratic model from the perspective of bipartite correlations in the ground state of the chain. Here we
report on the nature of correlations in the chain both for the nearest neighbor spins in the bulk and for a
small chain of 12 spins under open boundary condition. We should also mention in passing that the reason we also considered a small chain
of 12 spins in bilinear biquadratic model is to show the physical effects that can only be observed in the bulk
in this case. Looking at Fig. \ref{fig4} and Fig. \ref{fig5}, we can see that both first order transitions taking place
at TPs $\theta_{c2}=0.5\pi$ and $\theta_{c3}=1.25\pi$ are signaled by the discontinuous jumps in mutual
information and quantum discord, respectively. Moreover, exploring the correlations just for a chain of 12
spins is sufficient to detect these transitions. The difference between the behaviors of quantum discord and
mutual information at these points is that while mutual information first increases in a discontinuous fashion
at the TP $\theta_{c2}=0.5\pi$ and then again decreases at the TP $\theta_{c3}=1.25\pi$, quantum discord
behaves in the exact opposite way. When it comes to the second order transition occurring at the TP
$\theta_{c4}=1.75\pi$ between the trimerized and Haldane phases, Fig. \ref{fig4} and Fig. \ref{fig5} show that
it is not possible to pinpoint the TP in case of 12 spins since the curves of mutual information and quantum
discord are smooth without any sign of the transition. However, performing the same analysis for two spins
in the bulk, we see that a kink appears at $\theta=1.78\pi$, which in turn lets the second derivatives of both
mutual information and quantum discord to display a sharp maximum at the TP $\theta_{c3}=1.75\pi$. We emphasize
that this is different from the case of first order transitions, whose traces can be located regardless of the
size of the chain.

We now recall that, in the spin-1 bilinear biquadratic model, the point $\theta_{c1}=0.25\pi$ corresponds
to both the TP of the infinite order KT type transition and the $SU(3)$ symmetry point. In Fig. \ref{fig4},
we clearly observe a local minimum at this point, which would let one to conclude that the TP of the KT
transition, despite being an infinite order transition, can be detected through the behavior of mutual
information both for bulk and 12 spins. Nonetheless, we argue that what is signaled here is actually
the $SU(3)$ symmetry point of the model rather than the TP of the KT transition. Our argument is based
on what we have observed in case of the spin=1 XXZ model, that is, neither mutual information nor any
other studied measure are able to capture the KT transition point due to the analytical behavior of the
ground state energy and all of its derivatives. On the other hand, Fig. \ref{fig5} displays that a sharp
peak in quantum discord can be seen at this point, which we believe again spotlights the $SU(3)$ symmetry
point rather than the TP of the KT transition. In particular, our numerical treatment also reveals that
a sudden change emerges in the quantum discord at $\theta_{c1}=0.25\pi$ which has its roots in the change
of the optimizing basis in the definition of quantum discord. We stress that, in the spin-1 XXZ model,
the $SU(2)$ symmetry point has been also captured via discord through a sudden change, which supports
our argument that what we in fact observe here in the bilinear biquadratic model is the effect of the
$SU(3)$ symmetry point and not the KT transition happening at the same point. Also, we recall that
the measures for the 12 spins chain were unable to signal the second order TP at $\theta_{c4} = 1.75\pi$,
so it is quite unlikely they would detect a TP of infinite order. Finally, we point out that
the special points of the spin-1 bilinear biquadratic model, namely $\theta=0.1024\pi$ corresponding
to the AKLT model and $\theta=1.5\pi$, where the model has an exact solution with the Bethe ansatz
method, can be seen to be shown respectively in Fig. \ref{fig4} and Fig. \ref{fig5} through the extrema
of the mutual information and quantum discord.

As for the behavior of quantum coherence measures in case of the bilinear biquadratic model, even though we
do  not explicitly present our results here for the purposes of brevity and convenience, we have performed
an analysis similarly to the case of the XXZ model. We have observed that all three coherence measures are
able to capture the SU(3) symmetry point and the quantum phase transitions when two-spin coherence is studied.
On the other hand, we have seen for a single spin that $\sigma_x$-coherence can only detect the first order
phase transitions while the remaning two coherence measures vanish due to the fact that their density
matrices are diagonal in $S_z$ basis. This is in accord with the results of Ref. \cite{legeza} where the single site
entropy has been studied for the same model.

\section{Conclusion}

In summary, we have investigated the quantum mutual information, quantum discord and quantum
coherence in the ground states of spin-1 XXZ and bilinear biquadratic chains for two nearest
neighbor spins in the bulk. On one hand, our study has enabled us to draw conclusions regarding the
relation of the behavior of quantum correlations and coherence to the quantum phase transitions in these
models. On the other hand, we have established a link between the particular symmetry points of the
studied spin-1 Heisenberg chains and the considered correlation and coherence measures.

In particular, we have seen that neither the total and quantum correlations, as quantified by mutual
information and quantum discord respectively, nor the coherence measures have been able to
capture the TP of the infinite order KT type transition occurring in the XXZ model. However, they were
all able to locate the Ising type second order transition. In case of the $SU(2)$ symmetry
point, whereas we have observed that both of the correlation measures can detect it, $K$-coherence based
on Wigner-Yanase skew information is the unique coherence measure in our study which is able to signal
this symmetry point for a pair of nearest neighbor spins in the bulk. Furthermore, we have shown that
even for a single spin in the bulk, $K$-coherence can identify the $SU(2)$ symmetry and the Ising
transition in the spin-1 XXZ model.

Moreover, we have performed a similar analysis for the spin-1 bilinear biquadratic chain. Here, the mutual
information and quantum discord have signaled the point $\theta=0.25\pi$, which corresponds both to the TP
of the infinite order KT transition and the $SU(3)$ symmetry point. Based on our findings regarding the KT
transition and $SU(2)$ symmetry point in the XXZ chain, we have argued that what might actually be observed
through the measures is a consequence of the $SU(3)$ symmetry rather than the effect of the KT transition
occurring at the same point. Our argument is supported by the fact that quantum discord displays sudden changes
due to the discontinuous change of the optimizing basis in its definition at the symmetry points in both
models. Also, the density matrix elements of the nearest neighbor spins and all of their derivatives are
analytical at the KT transition points, thus the sudden changes in quantum discord are likely to have their
roots in the symmetries. We emphasize that the sudden change of quantum discord at the symmetry points is
fundamentally different from the behavior of quantum mutual information, which shows a local extremum at
these spots. Finally, we have pointed out the necessity of studying the correlations in the bulk, as opposed
to just a small chain of 12 spins, to be able to spotlight the second order transition in the chain.

\begin{acknowledgments}
ALM acknowledges the financial support from the Foundation for Development of UNESP (FUNDUNESP). GK is supported
by the S\~{a}o Paulo Research Foundation (FAPESP) under the grant numbers 2012/18558-5 and 2014/20941-7, FFF under the
grant number 2015/05581-7 and B\c{C} under the grant number 2014/21792-5. FFF is also supported by the National
Counsel of Technological and Scientific Development (CNPq) under the grant number 474592/2013-8 and by the National
Institute for Science and Technology of Quantum Information (INCT-IQ) under the process number 2008/57856-6. ROV
acknowledges the support from the Minas Gerais Research Foundation (FAPEMIG), CNPq and INCT-IQ.
\end{acknowledgments}

\end{document}